# kV-Class Lateral NiOx/GaN Super-Heterojunction Diode via Ammonia Molecular Beam Epitaxy (NH3-MBE)

Yizheng Liu[1,a)], Zachary J. Biegler[1], Ashley E. Wissel-Garcia[1], James S. Speck[1,a)], and Sriram Krishnamoorthy[1,a)]

[1]*Materials Department, University of California Santa Barbara, Santa Barbara CA 93106, USA*

a) *Author(s) to whom correspondence should be addressed. Electronic mail: yizhengliu@ucsb.edu, speck@ucsb.edu, sriramkrishnamoorthy@ucsb.edu*

***Abstract***: This work reports the demonstration of lateral p-NiOx/p-GaN/n-GaN-based super-heterojunction (SHJ) diodes using p-GaN with additional sputtered p-type nickel oxide (NiOx) layers to realize charge-balanced structures. The heterojunction diode capacitance-voltage (C-V) model is applied to extract effective the acceptor concentration from the p-NiOx. Net donor and acceptor concentration in n-GaN and p-GaN are extracted by using metal-oxide-semiconductor (MOS) test structures. The fabricated p-NiOx/p-GaN/n-GaN SHJ diodes with charge-balanced region between anode and cathode exhibit a forward on-state current density of 10-30 mA/mm across an anode-to-cathode distance ($L_{AC}$) from 16 μm to 80 μm. The SHJ diodes show rectifying behavior with a maximum on/off ratio of $10^9$ and a low reverse leakage density. The highest breakdown voltage achieved for the SHJ diodes is ~2.8 kV with reverse leakage density of $10^{-4}$ mA/mm at ~80% of device's catastrophic breakdown voltage. The SHJ diodes across all types of dimensions exhibit significant breakdown voltage improvements (~6× on average) with ultra-low reverse leakage current compared to corresponding reference structures without a charge-balanced extension, clearly demonstrating the superjunction effect for devices fabricated on GaN epitaxial layer with ~$10^{17}$ cm$^{-3}$ electron density.

Gallium nitride (GaN) and silicon carbide (SiC)-based wide-bandgap (WBG) power devices are increasingly adopted in low/medium voltage (600 V-1.7 kV) applications for electric vehicles (EVs), industrial motor drives, and renewable energy power processing. The performance limit of unipolar devices is constrained by the inherent trade-off between differential specific on-resistance ($R_{on,sp}$) and breakdown voltage ($V_{br}$) of power switches[1,2]. Charge-balanced superjunction devices offer a pathway to surpass this unipolar power figure of merit (PFOM), and break the trade-off between $R_{on,sp}$ and $V_{br}$, allowing devices to have high forward conduction via highly doped epilayers while maintaining improved breakdown voltage with a flattened electric field profile that is realized through careful charge-balancing between intentionally acceptor and donor doped regions[3–6]. Extensive research in WBG material-based superjunction devices, particularly in GaN, has been undertaken in recent years in forms of AlGaN/GaN single and multi-channel two dimensional electron gas (2DEG) charge-balanced with p-GaN, accomplishing breakdown voltages of ~8.85 kV and even beyond 10 kV[7–11]. In these reported lateral GaN super-heterojunction diodes, all the devices' active region epilayers are grown by metalorganic vapor chemical deposition (MOCVD), and charge balancing of these layers is accomplished as-deposited or via subsequent plasma-etching. The reverse leakage current density for these fabricated devices under large reverse bias varies significantly. In Xiao, *et al*'s AlGaN/GaN multi-channel 2DEG charge-balanced Schottky barrier diode (SBD) with p-GaN reduced surface electric field (RESURF) region, the reverse leakage current density is >$10^{-2}$ mA/mm reverse bias > 2.5 kV[10]. Han, *et al*'s AlGaN/GaN single channel 2DEG super-heterojunction SBDs can accomplish relatively low reverse leakage at $10^{-5}$-$10^{-3}$ mA/mm by introducing angled ion implantation to the superjunction devices' sidewall to eliminate the potential leakage pathway[7,9,11].

Recently, sputtered nickel oxide ($NiO_x$) has been widely used as an alternative p-type material for fabrication of heterojunction-based charge-balanced devices[6,12–15]. The conductivity of p-$NiO_x$ can be tuned during sputtering by controlling the oxygen-to-argon ratio ($O_2$/Ar)[15–17]. Literature results indicated that excess $O_2$ during sputter deposition can enhance the conductivity in the $NiO_x$, which is possibly attributed to Ni vacancy-mediated transport[18,19] in $NiO_x$. The equivalent acceptor concentration in sputtered $NiO_x$ can be extracted by using the heterojunction diode capacitance-voltage model below the diode's dispersion frequency[6,14,20], enabling this material to form charge-balanced superjunciton structure with various n-type WBG semiconductors[6,12,14,20,21].

This work presents the fabrication of $NiO_x$/GaN super-heterojunction (SHJ) lateral diodes by incorporating sputtered p-$NiO_x$ in combination with a thin p-GaN layer to accomplish post-growth charge-balance with ammonia molecular beam epitaxy ($NH_3$-MBE) grown n-GaN. The fabricated SHJ diodes with both p-GaN and p-$NiO_x$ charge-balancing layers exhibit a forward on-state current density of 10-30 mA/mm across an anode-to-cathode distance ($L_{AC}$) of 16-80 μm, and low reverse leakage current density at $10^{-6}$-$10^{-4}$ mA/mm with a high rectifying ratio ($J_{on}$/$J_{off}$) up to ~$10^9$. Breakdown voltages of the SHJ diodes are 0.94-2.8 kV with increasing $L_{AC}$, showing up to 6× on average breakdown voltage improvement compared to the SHJ reference diode structures without charge-balanced regions, indicating the beneficial effect of charge-balancing.

To accomplish accurate charge-balance, it is essential to characterize the net apparent charge concentrations in the n-GaN and p-GaN layers, as well as in the sputtered p-$NiO_x$. The n-GaN doping density characterization test structure, as shown in **Fig. 1(a)**, consisted of a $NH_3$-MBE grown unintentionally doped (UID) buffer layer (~200 nm) directly on top of a Fe-doped semi-insulating GaN-on-sapphire template, and a n-GaN (Si-doped) active region (~300 nm) capped by a ~100 nm of UID GaN. A metal-oxide-semiconductor (MOS) C-V structure was adopted to characterize the net charge density in n-GaN by using atomic layer deposition (ALD) grown silicon oxide ($SiO_2$~50 nm) underneath a Ni/Au (50/150 nm) metal stack. The contact to n-GaN was realized by an electron-beam evaporated Ti/Al/Ni/Au (30/120/30/50 nm) metal stack that was subsequently annealed at 820 °C in an ambient nitrogen ($N_2$) environment. The p-GaN doping density MOS-CV characterization test structure, as shown in **Fig. 1(b)**, was also grown on similar GaN template with a ~300 nm buffer region followed by a ~300 nm Mg-doped active region and capped with a 10~20 nm p$^{++}$ GaN contact layer. For characterization of p-GaN net acceptor concentration, Ni/Au (50/150 nm) was used as the metal stack on top of a sputtered SiO$_2$ (~50 nm), and large area Pd/Au (50/150 nm) stack served as Ohmic contact. Using MOS structures ensures low reverse leakage and reduced conduction loss under reverse bias for reliable C-V measurements.

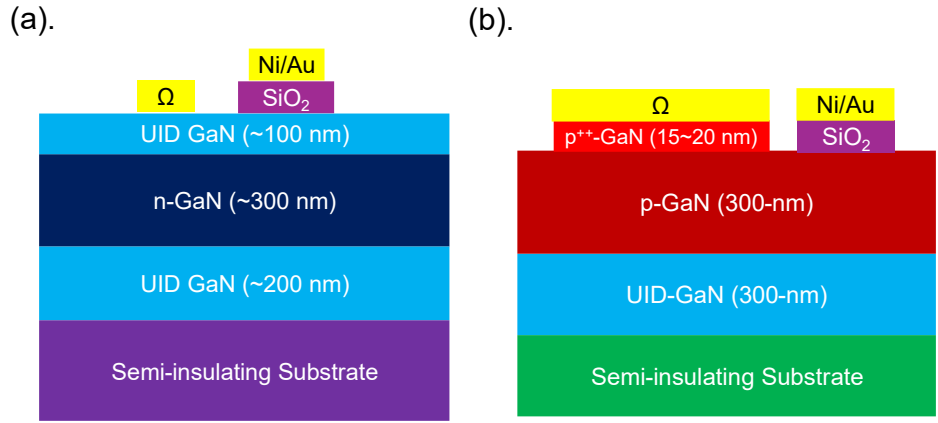

FIG.1. (a) MOS C-V structure of n-GaN SHJ epilayer. (b) MOS C-V structure of p-GaN doping reference epilayer with p$^{++}$ Ohmic contact layer.

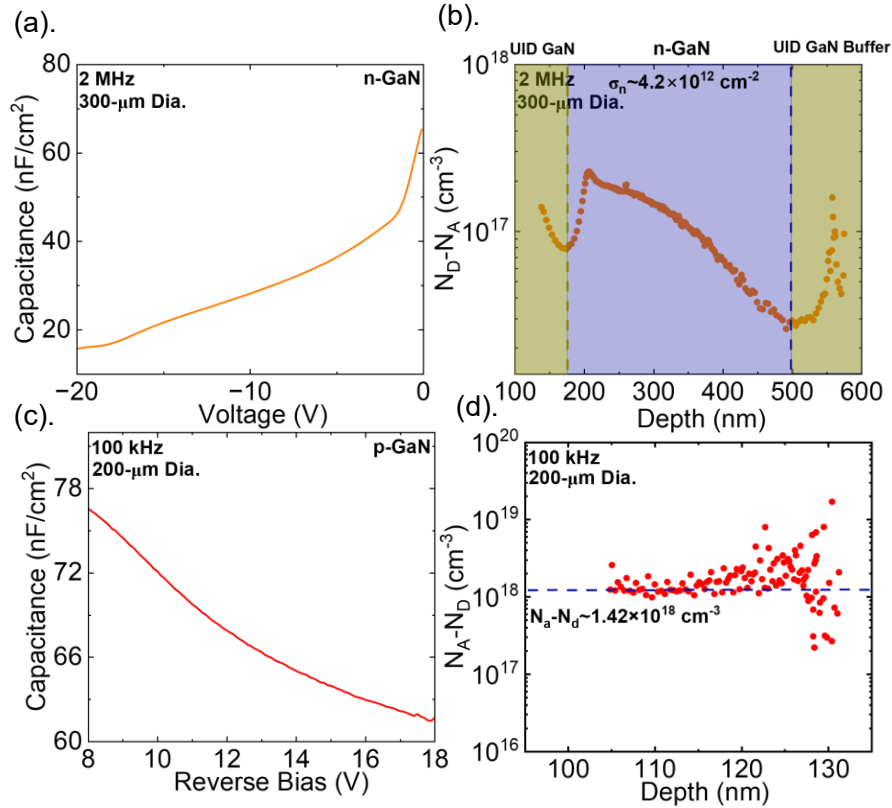

FIG.2. (a) C-V characteristics of MOS n-GaN SHJ epilayer at 2 MHz. (b) Net donor concentration vs. depth profile and a total charge density of 4.2 x 10$^{12}$ cm$^{-2}$ in the n-GaN and UID-GaN drift region. (c) C-V characteristics of MOS p-GaN SHJ epilayer at 100 kHz. (d) Net acceptor concentration vs. depth profile and a bulk extracted acceptor concentration at $1.42 \times 10^{18}$ cm$^{-3}$.

The total n-GaN and UID GaN net two-dimensional (2D) sheet ionized impurity density ($\sigma_n$) is integrated to be ~$4.2 \times 10^{12}$ cm$^{-2}$ at 2 MHz as shown in **Fig. 2(a) and 2(b)**, closely matching the sheet electron concentration at ~$4.9 \times 10^{12}$ cm$^{-2}$ from room temperature lithographically defined Van Der Pauw Hall effect measurements with an electron mobility of ~501 cm$^2$/V·s at an bulk net apparent charge density of ~$2.5 \times 10^{17}$ cm$^{-3}$. The slight reduction of the integrated sheet charge density from C-V measurements is possibly attributed to the incomplete depletion of the UID buffer region near the GaN-on-sapphire template substrate interface, indicating the presence of an interfacial spike carrier density caused by a large concentration of impurities at the regrowth interface[22,23]. Similarly, the p-GaN C-V characteristics at 100 kHz are shown in **Fig. 2(c)** with a net acceptor concentration ($N_A$-$N_D$) extracted at $1.42 \times 10^{18}$ cm$^{-3}$, as shown in **Fig. 2(d)**

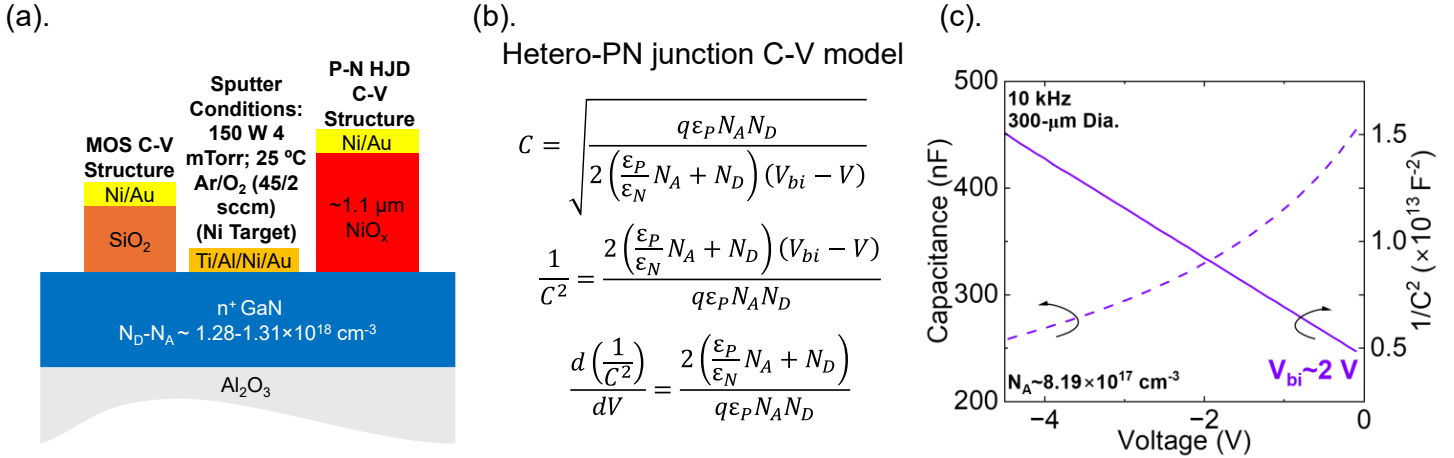

FIG.3. (a) MOS C-V and P-N diode structures for p-NiO$_x$ acceptor concentration ($N_A$) extraction. (b) Heterojunction P-N diode model for $N_a$ extraction in p-NiO$_x$. (c) C-V characteristics of p-NiO$_x$/GaN heterojunction at 10 kHz and 1/C$^2$ vs. voltage characteristics.

To extract the acceptor concentration ($N_A$) in sputtered p-NiO$_x$, a hetero-PN junction is fabricated on an n$^+$ GaN-on-sapphire template in **Fig. 3(a)** along with a Schottky MOS C-V structure used for net donor concentration determination in n$^+$ GaN. Using a highly doped GaN template ensures most of depletion happens in the thick p$^-$ NiO under reverse bias[24], and MOS C-V ensures suitably low conduction loss under reverse bias in the Schottky-based junction for reliable doping concentration (1.28~$1.31 \times 10^{18}$ cm$^{-3}$) extraction. The p-NiO$_x$ is sputtered via pure metallic Ni target with 45/2 sccm (Ar/O$_2$) with an RF power of 150 W and chamber pressure of 4 mTorr at room temperature[16,17]. By fitting the hetero-PN junction model in **Fig. 3(b)** using the C-V characteristics in **Fig. 3(c)** at 10 kHz[24], the $N_A$ in sputtered NiO$_x$ is extracted to be ~$8.19 \times 10^{17}$ cm$^{-3}$ with relative permittivity $\varepsilon_{r(NiO)}$ = 11.9[24] for NiO$_x$ and $\varepsilon_{r(GaN)}$ = 8.9 for GaN[1].

The epitaxial structure of the super-heterojunction (SHJ) diode is shown in the dashed box region in **Fig. 4(a)**. After the superjunction GaN epilayer growth using the same condition as the charge-balance for test structures (**Fig. 1(a) and 1(b)**), the device was fabricated starting with the mesa-isolation of the epitaxial GaN region by reactive ion etching (RIE) using a $BCl_3/Cl_2$ plasma at 100 W down to the Fe-doped semi-insulating substrates interface. Further patterned dry etching was carried out to remove the p-GaN (~20 nm) and UID GaN (~100 nm) to reach n-GaN active region for the Ohmic contact (cathode) formation. Immediately after dry etching, the wafers were rinsed in de-ionized water and were later submerged into a piranha etch solution ($H_2SO_4:H_2O_2 = 3:1$, Pure Strip) for 60 minutes to remove organic etching residue from chlorinated photoresist. A 3-minute buffered-oxide etch rinse was then performed to remove the potential oxide/hydroxide on sample surfaces from the piranha etch solution. After the oxide removal, the etched wafer was transferred to a heated 25% concentrated tetramethylammonium hydroxide (TMAH) solution at 50 °C for 5 minutes to remove potential sidewall dry etching damage from the $BCl_3/Cl_2$ plasma[25]. Finally, a 15-minute rapid thermal anneal (RTA) at 275 °C in the nitrogen ($N_2$) atmosphere was applied to reduce reverse leakage current on dry etch damaged GaN before cathode metallization[21,26]. The Ti/Al/Ni/Au (30/120/30/50 nm) metal stack was deposited on the n-GaN via e-beam evaporation and RTA annealed at 820 °C in an $N_2$ ambient for 30 seconds to form the Ohmic cathode of the SHJ diode. To complete the charge-balance with n-type ionized impurities, the p-$NiO_x$ extension region was sputtered on the planar p-GaN surface and device sidewall using identical deposition condition mentioned earlier in **Fig. 3(a) and 3(c)**. The p-type region and cathode separation distance ($L_{PC}$) was kept at 3 µm. For practical fabrication of the SHJ diode, if a charge imbalance margin is kept below 15%, the required p-$NiO_x$ thickness ($t_{NiOx}$) range can be estimated to be 21.9~29.6 nm, and 25.2 nm for the theoretically perfect charge-balance condition using the following relation

$$0.85 < \frac{Q_n}{Q_p} < 1.15 \qquad (1)$$

$$0.85 < \frac{4.9\times10^{12}cm^{-2}-(1.42\times10^{18}cm^{-3})(20\times10^{-7}cm)}{(8.19\times10^{18}cm^{-3})t_{NiO_x}} < 1.15 \qquad (2)$$

The sputtered $NiO_x$ was then annealed in an $N_2$ ambient at 275 °C for 15 minutes to stabilize its acceptor concentration[21]. After finalizing the p⁻ $NiO_x$ charge-balance region, a p⁺⁺ $NiO_x$ contact layer was sputtered (150 W/4 mTorr) with $Ar/O_2$ flows at 8/10 sccm. Then a self-aligned Ni/Au (50/150 nm) anode Ohmic metal stack was deposited by e-beam evaporation on the device's sidewall and 2-µm extension into the planar p-type region from the mesa edge. The higher percentage of $O_2$ in the $Ar/O_2$ mixture induces increased conductivity in the p-type $NiO_x$ for contact improvement[24]. After the metal liftoff from heated N-methyl-2-pyrrolidone (NMP) solution, an epoxy-based negative photoresist (SU-8) was applied to passivate the SHJ diode outside the contact region to mitigate potential electric arcing under high reverse bias[27]. To demonstrate the superjunction effect, reference diodes without the charge-balanced extension (0% charge-balance) were also fabricated along with the SHJ diodes, simultaneously, as shown in **Fig. 4(a) and 4(b)**.

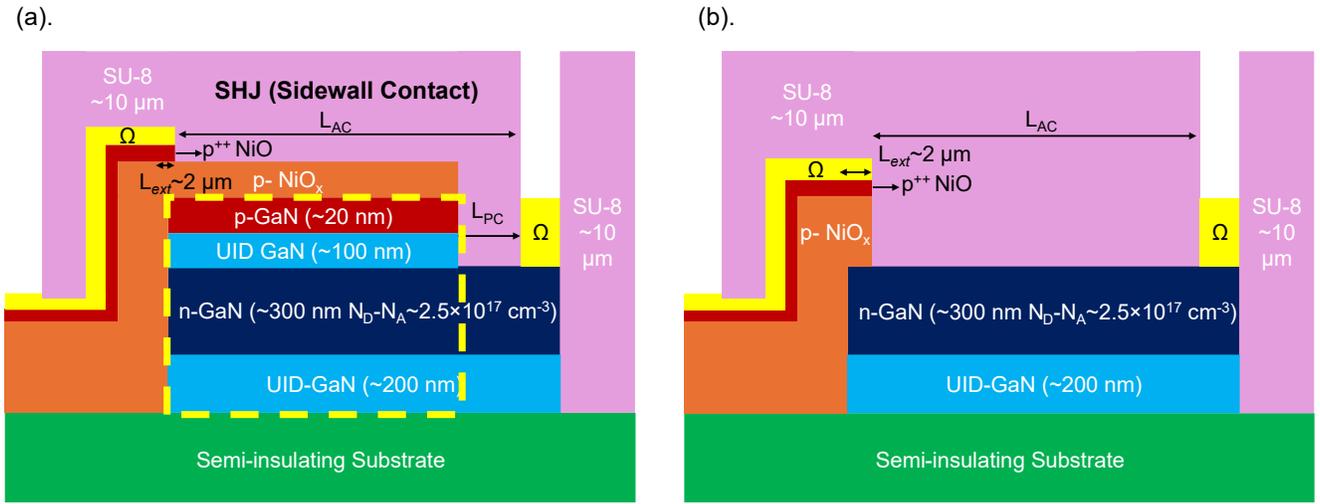

FIG.4. (a) NiO$_x$/GaN super-heterojunction diode passivated with SU-8. (b) NiO$_x$/GaN super-heterojunction diode reference structure (0% charge-balance) passivated with SU-8.

Ohmic contacts were obtained for both n-GaN and p-NiO$_x$ characterized by using circular transmission line measurements (CTLMs) test structures with electrode spacings ranging from 5 μm to 15 μm. The extracted sheet resistance from the n-GaN CTLM was at 5 kΩ/□ with a specific contact resistance of $1.21\times10^{-7}$ Ω·cm$^2$ for annealed Ti/Al/Ni/Au alloyed contact. The p-NiO$_x$ CTLM with a Ni/Au Ohmic stack gives a relatively high sheet resistance of 4.45 GΩ/□.

The super-heterojunction diodes with a 16 μm of anode-to-cathode distance (L$_{AC}$) exhibited rectifying behavior with an on-state current density reaching ~30 mA/mm at 5 V forward voltage as shown in **Fig. 5(a)**. The reference diode structure without the charge-balanced extension shows similar rectifying behavior but a higher current density at (~40 mA/mm) due to the absence of the vertical depletion region from p-type layers underneath the planar n-GaN top surface[20]. The current density of SHJ diodes decreases as L$_{AC}$ increases, as shown in **Fig. 5(b)-5(d)**. SHJ diodes with all four L$_{AC}$ exhibit rectifying behavior in their semi-log scale J-V characteristics in **Fig. 5(e)-5(h)**, with on-off ratios (J$_{on}$/J$_{off}$) of $10^6$-$10^9$ and reverse leakage current densities of $10^{-6}$-$10^{-4}$ mA/mm at -5 V on dislocated GaN-on-sapphire template (threading dislocation density ~$10^8$ cm$^{-2}$).

The SU-8 passivated SHJ diodes with 16 μm L$_{AC}$ exhibited catastrophic breakdown at 350-940 V with ultra-low reverse leakage current density at ~$10^{-5}$ mA/mm at 80% of devices' breakdown voltage, demonstrating significant breakdown voltage improvements compared to the reference structures, as shown in **Fig. 6(a)**. The breakdown voltages for p-NiO$_x$/p-GaN/n-GaN SHJ diodes with L$_{AC}$ at 25, 50, and 80 μm were 655-765 V, 1-1.5 kV, and ~2.8 kV, respectively as shown in **Fig. 6(b)-6(d)**, showing substantial device performance improvement compared to their corresponding reference diode structures without charge-balanced extensions. The majority of the SHJ breakdown characteristics showed ultra-low reverse leakage density at $10^{-6} \sim 10^{-4}$ mA/mm under elevated reverse bias at diodes' 80% breakdown voltages for devices fabricated on highly dislocated GaN-on-sapphire templates, demonstrating comparable performance in terms of reverse leakage current density compared to current state-of-the-art AlGaN/GaN-based superjunction devices[7–10]. The highest breakdown voltage of the SHJ diodes was accomplished at ~2.8 kV on device with anode-to-cathode distance at 80 μm on the n-GaN active region epilayer with > 1×10$^{17}$ cm$^{-3}$ apparent charge density, clearly demonstrating the superjunction effect compared to reference structures without charge-balanced regions for the NiO$_x$/GaN SHJ diodes in accomplishing high breakdown voltage without compromising the device's forward on-state conduction performance. To withstand an equivalent blocking voltage, it would otherwise require a conventional one-dimensional diode to have ~1×10$^{15}$ cm$^{-3}$ net n-type doping level over 80 μm, which is challenging to accomplish in current GaN epitaxy, and degrades device's on-state resistance.

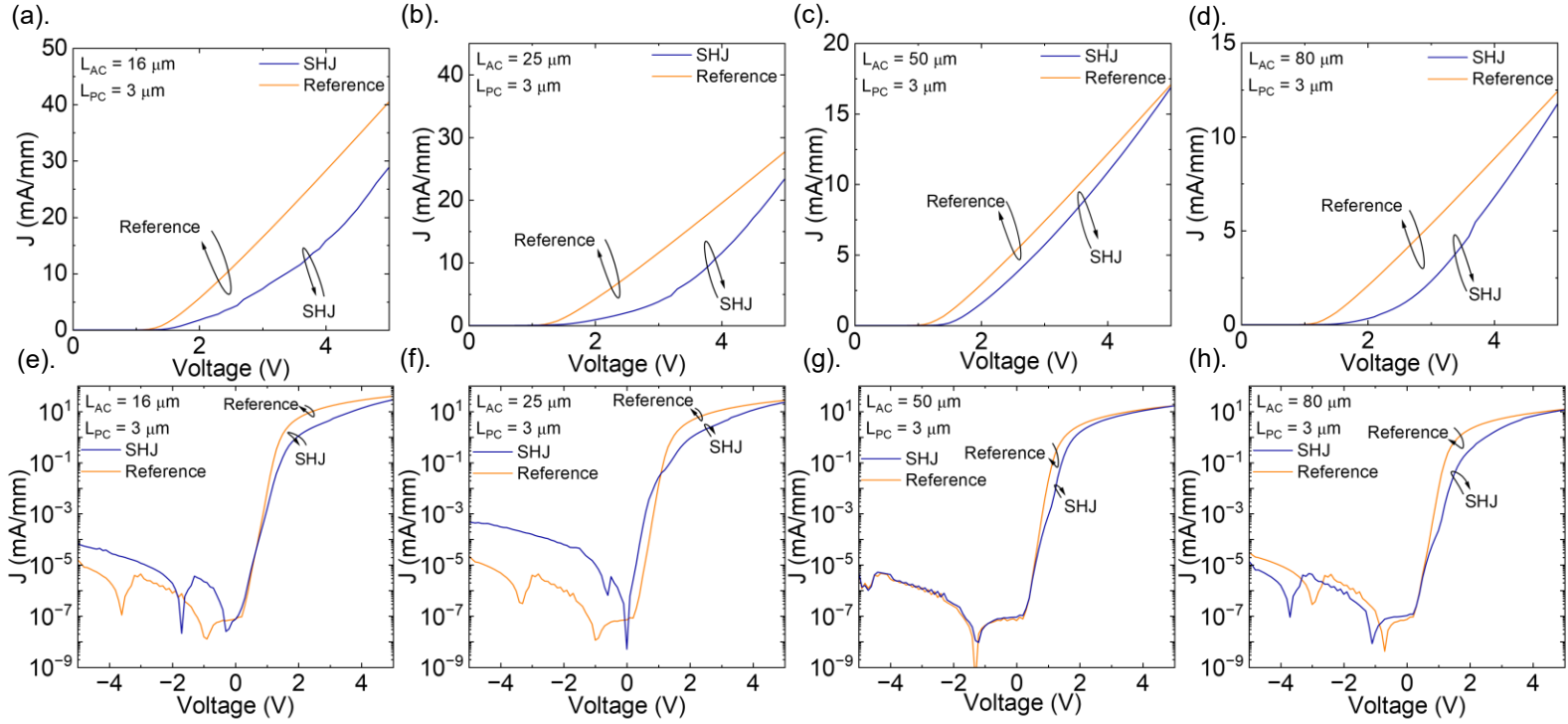

FIG.5. (a)-(d) Linear forward current density vs. voltage characteristics of the NiO$_x$/GaN SHJ diodes and reference structures with anode-to-cathode distance (L$_{AC}$) from 16 μm to 80 μm. (e)-(h) Semi-log scale current density vs. voltage characteristics of the NiO$_x$/GaN SHJ diodes and reference structures with anode-to-cathode distance (L$_{AC}$) from 16 μm to 80 μm.

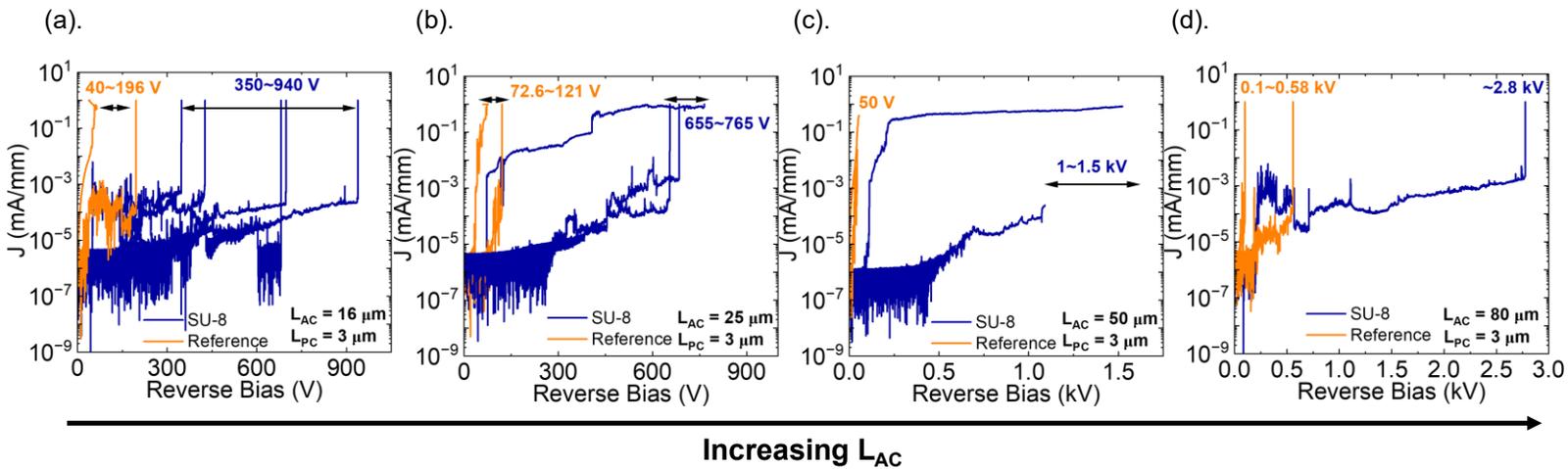

FIG.6. Reverse breakdown characteristics of NiO$_x$/GaN SHJ diode and reference structures with L$_{AC}$ at (a) 16 μm, (b) 25 μm, (c) 50 μm, and (d) 80 μm.

In summary, this work demonstrates GaN-based super-heterojunction diodes by using NH$_3$-MBE-grown GaN epitaxial layers and shows the heterogenous integration of p-NiO$_x$ with GaN for achieving charge-balance. In comparison with the reference structure, the highest breakdown voltage of NiO$_x$/GaN SHJ is at ~2.8 kV, showing up to 6-fold improvement compared to reference structures without charge-balanced regions and clearly demonstrating the superjunction effect for devices with reasonably conductive on-state performance, that are fabricated on >1×10$^{17}$ cm$^{-3}$ apparent doping density GaN epilayer. The reverse leakage current density at 80% of SHJ diode's breakdown voltage is at 10$^{-6}$-10$^{-4}$ mA/mm, which is among the lowest reverse leakage current density for GaN superjunction diodes[7–10].

**ACKNOWLEDGMENTS**

The authors acknowledge funding from the U.S. Department of Energy (DOE) ARPA-E OPEN 2021 program (DE-AR0001591). A portion of this work was performed at the UCSB Nanofabrication Facility, an open access laboratory.

**DATA AVAILABILITY**

The data that supports the findings of this study are available from the corresponding authors upon reasonable request.